\documentclass[aps,prd,nofootinbib,groupedaddress,twocolumn]{revtex4}

\usepackage{url}
\usepackage{graphicx}
\usepackage[utf8]{inputenc}
\usepackage{amsmath,amssymb,amstext,amssymb,amsfonts,amsthm}
\usepackage{hyperref}
\usepackage[margin=1.0in,papersize={8.5in,11in}]{geometry}

\begin{document}

\title{General Relativistic Corrections to the Weak Lensing Convergence Power Spectrum}

\author{John T. Giblin, Jr${}^{1,2}$}
\author{James B. Mertens${}^{2}$}
\author{Glenn D. Starkman${}^{2}$}
\author{Andrew R. Zentner${}^{3}$}

\affiliation{${}^1$Department of Physics, Kenyon College, 201 N College Rd, Gambier, OH 43022}
\affiliation{${}^2$CERCA/ISO, Department of Physics, Case Western Reserve University, 10900 Euclid Avenue, Cleveland, OH 44106}
\affiliation{${}^3$Department of Physics and Astronomy, University of Pittsburgh, 100 Allen Hall, 3941 O\'Hara St., Pittsburgh PA 15260}

\begin{abstract}
We compute the weak lensing convergence power spectrum, $C^{\kappa\kappa}(\theta)$,
in a dust-filled universe using fully non-linear general relativistic simulations.
The spectrum is then compared to more standard, approximate calculations by
computing the Bardeen (Newtonian) potentials in linearized gravity and utilizing
the Born approximation. We find corrections to the angular power spectrum amplitude
of order ten percent at very large angular scales, $\ell \sim 2-3$, and percent-level
corrections at intermediate angular scales of $\ell \sim 20-30$.
\end{abstract}

\maketitle

\section{Introduction}

Weak lensing calculations rely on a number of assumptions in order to improve
tractability of models. These include physical assumptions, such as the Born
approximation, where physical arguments are used to justify neglecting
sub-dominant effects. Further assumptions are made when modeling the
gravitational physics of lensed systems, in particular the assumption that a
linearized gravitational model provides a sufficiently accurate description of
the dynamics of the evolution of the Universe as well as the geodesic equations
describing propagation of light. Here, we explore the impact of these
assumptions on weak lensing convergence calculations by comparing standard,
commonly used calculations to a fully general relativistic treatment
of the problem. We find these approximations are accurate only to within a few
percent on large angular scales. The relative magnitude of corrections is found
to increase on larger angular scales, and lessen on smaller angular scales.

Such observations of weak lensing convergence power spectra are among the
primary science goals of the ongoing Dark Energy Survey (DES) and Hyper
Suprime-Cam (HSC) Subaru Strategic Survey, as well as the forthcoming surveys
of the Large Synoptic Survey Telescope (LSST), the Euclid mission, and
Wide-Field InfraRed Survey Telescope (WFIRST). The primary driver behind
these measurements is their potential to use lensing power spectra to
constrain cosmology, particularly the cause of cosmic acceleration
\cite{1201.2434} and neutrino mass (for example,
Refs.~\cite{1209.1043,1405.6205,1310.0037}). Observationally viable models of
dark energy and values of neutrino masses induce only subtle alterations to
lensing power spectra on the order of a few percent. Consequently, it is of
critical importance to produce theoretical predictions for lensing power
spectra that are both very precise and very accurate so that the data are not
misinterpreted (e.g., Refs.~\cite{astro-ph/0412142,1111.0052,1405.6205}). This
continues to be one of the challenges to the exploitation of weak lensing
observations for cosmological analyses.

Carefully examining the physical and perturbative approximations made in the
context of lensing measurements requires a number of subtle considerations in
order to compare to a fully relativistic treatment, ranging from the
gauge-dependent nature of variables used in calculations to the particular way
in which averaged quantities are utilized. We attempt to remain self-consistent
in our treatment of this problem, and to explicitly define the quantities we
consider and approximations we use. We begin by defining angular
diameter distances and convergence in terms of optical scalars, and describe
the 3+1 framework we use to numerically integrate Einstein's equations. We
then compute the Bardeen (Newtonian) gravitational potentials, use these
potentials to obtain the weak lensing convergence field in an approximate
setting, and compare the two models.

Past literature has explored the magnitude of corrections to observables
due to commonly made assumptions \cite{astro-ph/0512159,1205.3366},
speculating that contributions from nonlinear gravitational effects can lead
to approximately percent-level corrections, contingent upon on the specific
statistical measure being studied
\cite{1209.3142,1207.2109,1402.4350,1612.00852}. Here, we perform the first
such study in a fully relativistic setting, utilizing simulations of a universe
containing a cosmologically-motivated spectrum of density fluctuations in a
perfect, pressureless ``dust'' fluid.
We find percent-level corrections to the convergence power spectrum at
$\ell \sim 10-20$. The relative importance of corrections is found to
increase at smaller $\ell$, becoming of order ten percent at $\ell$ of a
few. At higher $\ell$, the relative importance of relativistic corrections
is found to decrease -- although perhaps a physical effect, this may also
be a consequence of the spectrum of perturbations that we used, which
contains only long-wavelength modes.

We begin by briefly detailing the methods we use to perform the
fully relativistic calculation and the approximate calculations we compare
to. We then present a quantitative comparison of simulated quantities using
the two methods. We begin in Section~\ref{sec:Formalism} by describing the
different formalisms we utilize to perform numerical calculations. In
Section~\ref{sec:Initial-Conditions} we describe initial conditions for
the toy universes we utilize, and in Section~\ref{sec:Results} we detail
results from numerical simulations.

\section{Formalism\label{sec:Formalism}}

\subsection{\label{subsec:raytracing}Raytracing in the BSSNOK formulation}

The field of numerical relativity has evolved over the past several
decades to become a standard numerical tool in contemporary physics. The field
has progressed to the point where it can model physics ranging from systems of
strongly gravitating compact objects in a fully relativistic, cosmological setting
\cite{Bentivegna:2013jta,Yoo:2014boa,1611.09275}, to the dynamics of perfect fluids
as they interact on cosmological length scales
\cite{Giblin:2015vwq,Bentivegna:2015flc,1704.04307}.
The BSSNOK formulation is a commonly used numerical scheme that has been
demonstrated capable of modeling such systems with a high degree of accuracy,
and importantly, numerical stability \cite{Nakamura:1987zz,Shibata:1995we,Baumgarte:1998te}.

The BSSNOK system of equations is a 3+1 conformal decomposition of
the Einstein field equations. In this language, the line element is 
\begin{equation}
ds^{2}=-\alpha^{2}dt^{2}+e^{4\phi}\bar{\gamma}_{ij}\left(dx^{i}+\beta^{i}dt\right)\left(dx^{j}+\beta^{j}dt\right)\,,
\end{equation}
where $e^{4\phi}\bar{\gamma}_{ij}$ is the spatial metric, and $\bar{\gamma}_{ij}$
is a unit-determinant matrix. The parameters $\alpha$ and $\beta^{i}$ are
respectively known as the lapse and shift. Einstein's field equations can be
written in terms of these variables as a system of first-order dynamical
equations,
\begin{align}
\partial_{t}\phi= & -\frac{1}{6}\alpha K+\beta^{i}\partial_{i}\phi+\frac{1}{6}\partial_{i}\beta^{i}\\
\partial_{t}\bar{\gamma}_{ij}= & -2\alpha\bar{A}_{ij}+\beta^{k}\partial_{k}\bar{\gamma}_{ij}+\bar{\gamma}_{ik}\partial_{j}\beta^{k}\nonumber\\
 & +\bar{\gamma}_{kj}\partial_{i}\beta^{k}-\frac{2}{3}\bar{\gamma}_{ij}\partial_{k}\beta^{k}\\
\partial_{t}K= & -\gamma^{ij}D_{j}D_{i}\alpha+\alpha(\bar{A}_{ij}\bar{A}^{ij}+\frac{1}{3}K^{2})\nonumber \\
 & +4\pi\alpha(\rho+S)+\beta^{i}\partial_{i}K\\
\partial_{t}\bar{A}_{ij}= & e^{-4\phi}(-(D_{i}D_{j}\alpha)+\alpha(R_{ij}-8\pi S_{ij}))^{TF}\nonumber \\
 & +\alpha(K\bar{A}_{ij}-2\bar{A}_{il}\bar{A}_{j}^{l}) +\beta^{k}\partial_{k}\bar{A}_{ij} \nonumber \\
 & +\bar{A}_{ik}\partial_{j}\beta^{k}+\bar{A}_{kj}\partial_{i}\beta^{k}-\frac{2}{3}\bar{A}_{ij}\partial_{k}\beta^{k}\,.
\end{align}
The lapse and shift are considered gauge variables, and may be freely
chosen. An additional auxiliary variable, a contraction of a conformal
Christoffel symbol, $\bar{\Gamma}^{i}=\bar{\gamma}^{jk}\bar{\Gamma}_{jk}^{i}$,
is evolved to improve numerical stability properties of the system
according to
\begin{align}
\partial_{t}\bar{\Gamma}^{i}= & -2\bar{A}^{ij}\partial_{j}\alpha+2\alpha\big(\bar{\Gamma}_{jk}^{i}\bar{A}^{jk}-\frac{2}{3}\bar{\gamma}^{ij}\partial_{j}K\nonumber \\
 & -8\pi\bar{\gamma}^{ij}S_{j}+6\bar{A}^{ij}\partial_{j}\phi\big) +\beta^{j}\partial_{j}\bar{\Gamma}^{i} -\bar{\Gamma}^{j}\partial_{j}\beta^{i}\nonumber\\
 & +\frac{2}{3}\bar{\Gamma}^{i}\partial_{j}\beta^{j}+\frac{1}{3}\bar{\gamma}^{li}\partial_{l}\partial_{j}\beta^{j}+\bar{\gamma}^{lj}\partial_{l}\partial_{j}\beta^{i}\,,
\end{align}
and is used when to computing the Ricci tensor and scalar.

Likewise, we can integrate the optical scalar equations \cite{Sachs:1961zz}
using the full framework of general relativity. Optical integration through
a spacetime involves tracking beam ``areas'' along photon geodesics. The
cosmological observable we compute here is the angular diameter distance,
$D_{A}\equiv\ell/\Omega$, for an object with some physical length $\ell$ that
subtends an angle $\Omega$ of an observer's sky. The optical scalar equations
are valid in the limit of infinitesimal beams, or in the limit that both
$\ell$ and $\Omega$ are small, although recent work may offer a way of working
around this limitation \cite{1706.09383}. The optical scalar equations assume
photons do not interact, or that the beam follows photon geodesics and neither
backreact nor interact with other matter in the universe. The optical scalar
equations are given by
\begin{equation}
\frac{{\rm d}^{2}}{{\rm d}\lambda^{2}}\ell  =\ell\left(\mathcal{R}-\sigma^{2}\right)
\end{equation}
and
\begin{equation}
\frac{{\rm d}}{{\rm d}\lambda}\left(\ell^{2}\sigma\right)  =\ell^{2}\mathcal{W}\,,
\end{equation}
for some affine parameter $\lambda$ along a photon path, beam area
$\ell$, shear rate $\sigma$, and Ricci and Weyl optical
scalars, $\mathcal{R}$ and $\mathcal{W}$. Further details about these
equations and our previous work numerically integrating these equations
can be found in \cite{1608.04403} and references therein.

As a final point of potential interest, we remark upon the computational
complexity of the scheme described above. In synchronous gauge where $\alpha = 1$, used
in this work, and indeed in the vast majority of gauges typically
used in numerical relativity, Einstein's equations are completely
local, so calculations are $\mathcal{O}(N)$ for some number $N$
of discretized elements of interest (grid points, particles, ...).
This is in contrast to the use of a nonlocal gauge, where calculations
typically scale as $\mathcal{O}(N\log N)$. This penalty is incurred
when using Newtonian gauge, commonplace in N-body simulations, and
is encountered in this work when we compute the Bardeen
potentials. The drawbacks of a fully relativistic calculation are
due in part to the increased number of algebraic calculations involved,
but perhaps more important is the need to resolve luminal propagation.
However, these penalties should also be incurred by any code wishing
to reliably integrate geodesics, resolve luminally propagating phenomena,
or resolve higher-order gravitational effects, even within a framework
of linearized gravity.

\subsection{Computing Convergence}
\label{subsec:convergence}

Convergence in weak lensing may be defined in terms of angular diameter
distances as
\begin{equation}
\kappa=\frac{\bar{D}_{A}-D_{A}}{\bar{D}_{A}}\,,\label{eq:gr_convergence}
\end{equation}
where $\bar{D}_{A}$ is the angular diameter distance as defined in
a pure-FLRW universe \cite{1508.07903}. Defined this way,
convergence is meaningful in a fully relativistic setting, ie. no
perturbative assumptions need to be made, and the above expression
reduces to expressions found in cosmological literature in a Newtonian
setting.

In a general relativistic setting, computing angular diameter distances, and
thus convergence as defined in Eq.~\ref{eq:gr_convergence}, requires
integration of the optical scalar equations as detailed in Sec.~\ref{subsec:raytracing}.
In Newtonian gauge (sometimes referred to as Poisson gauge or longitudinal
gauge; here we follow the conventions of \cite{Weinberg:2008zzc}), this task
is simplified after making several assumptions, both perturbative and physical
\cite{1601.02012,1612.00852,1402.4350}. Typically, perturbative assumptions
enter by modeling spacetime and matter dynamics within a linearized gravity
framework, while the physical assumptions include assuming the behavior of the
spacetime is sufficiently well-described by scalar degrees of freedom and
perfect fluid components. Using these assumptions, this expression for weak
lensing convergence can be written in terms of the Bardeen (Newtonian)
potential $\Phi$,
\begin{equation}
\kappa=\int\left(r_{s}-r\right)\frac{r}{r_{s}}\nabla_{\perp}^{2}\Phi dr\,.\label{eq:linear_convergence_expr}
\end{equation}
The gradient in this expression is transverse to the direction of
propagation, $\hat{n}$, thus can alternatively be written in terms
of a full Laplacian minus a component along this direction,
\begin{equation}
\nabla_{\perp}^{2}\Phi=\nabla^{2}\Phi-\partial_{\hat{n}}^{2}\Phi\,.\label{eq:transverse_grad_decomp}
\end{equation}
The full Laplacian may be evaluated using the Hamiltonian constraint
equation in Newtonian gauge linearized around an FLRW background in
the presence of a perfect fluid, or Poisson's equation for gravity,
\begin{equation}
\label{eq:newtonian_poisson}
\nabla^{2}\Phi=4\pi a^{2}\delta\rho+12\pi a^{2}(\bar{\rho}+\bar{p})\delta u\,.
\end{equation}
We will also compute the radial coordinate using the Born approximation,
\begin{equation}
r(z)=\int_{0}^{z}\frac{1}{H(z')}dz'\,.
\end{equation}
This is the only time we use the Born approximation---the Bardeen potential
and stress-energy quantities are evaluated along a true geodesic, computed
using the fully general relativistic expression. The derivative of the Bardeen
potential along the path of integration is also computed along this geodesic,
with photon redshift rescaled to a coordinate expression using the Born
approximation. The second derivative is then computed with respect to this
radial coordinate.

Often, the radial derivative and peculiar velocity contributions are
neglected entirely; we do not include these assumptions in our analysis.
The peculiar velocity contribution can be accounted for by adding a term
proportional to the fluid velocity components at the source and observer,
$\vec{v}_s$ and $\vec{v}_o$ along the line of sight $\hat n$ in Newtonian
gauge \cite{1411.6339}, $\kappa \rightarrow \kappa + \kappa_{v}$, where
\begin{equation}
\kappa_{v} = -\frac{a}{\dot{a} \int dz/H} (\vec{v}_s - \vec{v}_o ) \cdot \hat{n} + \vec{v}_s \cdot \hat{n}\,.
\end{equation}

The final approximate expression for convergence we seek to integrate is thus
\begin{multline}
\kappa = \int\left(r_{s}-r\right)\frac{r}{r_{s}} \big(4\pi a^{2}\delta\rho \\
 + 12\pi a^{2} (\bar{\rho}+\bar{p})\delta u -\partial_{\hat{n}}^{2}\Phi\big)dr + \kappa_{v}\,.\label{eq:phys_appx_convergence}
\end{multline}
For a final comparison, we compute this approximate expression,
Eq. \ref{eq:phys_appx_convergence}, and compare to Eq. \ref{eq:gr_convergence},
a fully general relativistic result.

As a final note, the precise magnitude of corrections will depend upon the
background cosmological parameters that are chosen. To this end, we note that
we compute $H(z)$, $a(z)$, and $\bar{D}_{A}(z)$ using a background cosmology
defined by the initial conditions in the simulation.

\subsection{Computing Bardeen potentials from a fully relativistic simulation\label{subsec:sync2newton}}

The Bardeen potentials, or Newtonian potentials $\Phi$ and $\Psi$,
may be computed from a known metric in an arbitrary gauge---here, we compute
them using the synchronous gauge (geodesic slicing) metric. We obtain the
Bardeen potentials by first performing a scalar-vector-tensor (SVT)
decomposition of the metric linearized around a homogeneous FLRW background.
We perform this decomposition following Weinberg \cite{Weinberg:2008zzc},
writing the metric as a background plus perturbation,
\begin{equation}
g_{\mu\nu}=\bar{g}_{\mu\nu}+h_{\mu\nu}.
\end{equation}
An ambiguity exists in this definition in that the choice of $\bar{g}_{\mu\nu}$
is arbitrary---any background metric will suffice---however the FLRW
metric is chosen in a cosmological setting as we expect the dynamics
of the spacetime to be well-described by such a background,
so $\bar{g}_{\mu\nu}={\rm diag}(-1,a^{2}\delta_{ij})$. Fluctuations
around this background are taken to be small so that quantities derived
from $h_{\mu\nu}$ may be raised and lowered using purely the background
metric with terms second-order in $h_{\mu\nu}$ dropped. The perturbed
metric is then decomposed as
\begin{widetext}
\begin{equation}
h_{\mu\nu}=\left(\begin{array}{cc}
h_{00} & h_{i0}\\
h_{0j} & h_{ij}
\end{array}\right)=\left(\begin{array}{cc}
-E & a\left(\partial_{i}F+G_{i}\right)\\
a\left(\partial_{j}F+G_{j}\right) & \,\,a^{2}\left(A\delta_{ij}+\partial_{i}\partial_{j}B+\partial_{i}C_{j}+\partial_{j}C_{i}+D_{ij}\right)
\end{array}\right)\,.
\end{equation}
\end{widetext}
The vector and scalar functions are transverse with respect to the
background metric, 
\begin{equation}
\partial_{i}C^{i}=\partial_{i}G^{i}=\partial_{i}D^{ij}=0\,,
\end{equation}
and the tensor perturbation is trace-free, $D_{i}^{i}=0$. Given that the
3+1 metric in synchronous gauge is written as
\begin{equation}
g_{\mu\nu}=\left(\begin{array}{cc}
-1 & 0\\
 0 & \gamma_{ij}
\end{array}\right)\,,
\end{equation}
we immediately see that in synchronous gauge, a number of potentials are
zero, $F = G_i = E = 0$. We can also see that the metric perturbations
and their time derivatives (which will be needed to compute the Bardeen
potentials) can be written in terms of BSSNOK variables as
\begin{align}
h_{ij} & =\gamma_{ij}-a^{2}\delta_{ij}\\
\dot{h}_{ij} & =-2K_{ij}-2a\dot{a}\delta_{ij}\\
\ddot{h}_{ij} & =-2\left(\dot{K}_{ij}+\left(\dot{a}^{2}+a\ddot{a}\right)\delta_{ij}\right)\,,
\end{align}
where $K_{ij}=e^{4\phi}\bar{A}_{ij}+\frac{1}{3}\bar{\gamma}_{ij}K$
is the extrinsic curvature, and its time derivative can be written
in terms of BSSNOK variables,
\begin{align}
\dot{K}_{ij} = & 4\dot{\phi}e^{4\phi}\left(\bar{A}_{ij}+\frac{1}{3}\bar{\gamma}_{ij}K\right) \nonumber \\
 & +e^{4\phi}\left(\dot{\bar{A}}_{ij}+\frac{1}{3}\dot{\bar{\gamma}}_{ij}K+\frac{1}{3}\bar{\gamma}_{ij}\dot{K}\right)\,.
\end{align}
From this, we can reconstruct the SVT scalar field A,
\begin{equation}
A=\frac{1}{2 a^2}\left(h_{ii}-\frac{1}{\nabla^{2}}\partial_{i}\partial_{j}h_{ij}\right)\,,
\end{equation}
and $B$, 
\begin{equation}
B=\frac{1}{\nabla^{2}}\left(\frac{h_{ii}}{a^2}-3A\right)\,,
\end{equation}
with time-derivatives of $B$ being computed using time derivatives of
$h_{ij}$. An ambiguity in this definition of $B$ allows for the addition of
an arbitrary time-dependent function. To address this, we note that we
specify the zero-mode of the inverse Laplacian to be zero. We numerically
solve these equations for $A$ and $B$ in Fourier space. The remaining
vector and tensor potentials may also be determined if desired, however we
do not do so here. At this point, we have enough information to construct
the Bardeen potentials. In terms of the synchronous gauge scalar potentials,
these are given by
\begin{align}
\Phi & =-\frac{a}{2}\left(2\dot{a}\dot{B}+a\ddot{B}\right)\nonumber \\
\Psi & =\frac{1}{2}\left(a\dot{a}\dot{B}-A\right)\,.
\end{align}
As mentioned before, there is one further minor ambiguity: the scale factor
$a$ can be chosen in several different ways. For example, it can be chosen to
correspond to the average conformal factor $\left\langle e^{2\phi}\right\rangle$
in a particular slicing, or the FLRW solution corresponding to this value computed
on the initial or final slices. For this work, we opt to choose a scale factor
that coincides with the scale factor on the initial surface, and that evolves
according to the standard matter-dominated Friedmann equations.

Converting density fluctuation amplitudes from synchronous gauge to Newtonian
should be performed as well,
\begin{equation}
\delta\rho^{N} =\delta\rho^{S}+\frac{a^{2}}{2}\dot{B}\dot{\bar{\rho}}\,,
\end{equation}
along with fluid 3-velocity velocity, $\delta u_i = \partial_i \delta u$,
\begin{equation}
\delta u^{N} =\delta u^{S}-\frac{a^{2}}{2}\dot{B}\,.\label{eq:matter_sync2newt}
\end{equation}
As a final note, although the Synchronous gauge metric is not uniquely determined
in terms of the Bardeen potentials, we do not transform variables from Newtonian
to synchronous, and thus do not encounter this issue.

\section{Initial Conditions\label{sec:Initial-Conditions}}

We set initial conditions by generating a random realization of a
cosmologically-motivated power spectrum, similar to past work \cite{1511.01106}.
As the initial conditions we use are intended to mimic an inhomogeneous
cosmology, we attempt to, at least approximately, match large-scale
matter density fluctuations. At large scales, the power spectrum of
density fluctuations is expected to scale as $P_{\delta}\propto k^{1}$,
and at small scales as $P_{\delta}\propto k^{-3}$. We choose a spectrum
that corresponds to these scalings,
\begin{equation}
P_{\delta\delta}=\frac{4}{3}P_{*}\frac{k/k_{*}}{1+(k/k_{*})^{4}/3}\times C(k,k_{c})\,,
\end{equation}
with $k_{*}$ the peak frequency and $P_{*}$ the amplitude of the
power spectrum. The function $C$ is included in order to introduce
a short-wavelength cutoff scale, $k_{c}$, to exclude small-scale
modes that are not well-resolved and can therefore lead to numerical
instability or inaccuracy. In practice, this means resolving all modes
by $\mathcal{O}(5-10)$ grid points or more on the initial surface.
We choose $C$ to be a logistic function,
\begin{equation}
C(k,k_{c})=\frac{1}{1+e^{10(k-k_{c})}}\,.
\end{equation}
We additionally choose the initial peak frequency to correspond to
a length scale of roughly 300 Mpc, and a power spectrum amplitude
that corresponds to a realistic RMS amplitude of the density. Although 8
Mpc scales are not well-resolved, we still use a power spectrum amplitude
that corresponds to a $\sigma_{8}$ value (RMS density fluctuation
amplitude smoothed on 8 Mpc scales) of $\sigma_{8}\sim0.8$. We simulate
half of a Hubble volume and include modes down to $k^{-1} = 1/40\,H^{-1} \sim 100$
Mpc. Smoothed on this scale, the expected RMS density amplitude is
$\sigma_{100}\sim0.07$ at the time of observation \cite{Lyth:2009zz}.
We choose our power spectrum amplitude such that the amplitude of the
conformal RMS density fluctuations, defined as
\begin{equation}
\sigma_{\rho}=\sqrt{\frac{\int dV\sqrt{\gamma}\left(\bar{\rho}-\rho\right)^{2}}{\int dV\sqrt{\gamma}}}\,,
\end{equation}
with the average density defined as
\begin{equation}
\bar{\rho}=\frac{\int dV\sqrt{\gamma}\rho}{\int dV\sqrt{\gamma}}\,,
\end{equation}
approximately coincides with this value of $\sigma_{100}$.

The metric and matter fields on the initial surface must satisfy the
Hamiltonian and momentum constraint equations. In terms of BSSN
variables, these are given by
\begin{multline}
\mathcal{H} = 0 = \bar{\gamma}^{ij}\bar{D}_{i}\bar{D}_{j}e^{\phi}-\frac{e^{\phi}}{8}\bar{R} \\
+\frac{e^{5\phi}}{8}\bar{A}^{ij}\bar{A}_{ij} -\frac{e^{5\phi}}{12}K^{2}+2\pi e^{5\phi}\rho
\end{multline}
and
\begin{equation}
\mathcal{M}^{i}  = 0 = \bar{D}_{j}\left(e^{5\phi}\bar{A}^{ij}\right)-\frac{2}{3}e^{6\phi}\bar{D}^{i}K-8\pi e^{10\phi}S^{i}\,.
\end{equation}
We choose $\bar{A}_{ij}=0$, and $\bar{\gamma}_{ij}=\delta_{ij}$,
imposing the restriction that the 3-metric be conformally flat on the initial slice. The
momentum constraint can be trivially solved by choosing the extrinsic
curvature $K$ to be constant and the momentum variable $S^{i}$ to
be zero, consistent with a fluid initially at rest. The Hamiltonian
constraint equation can be solved by specifying the remaining metric
components, and solving for the corresponding density. For the conformally
flat metric we have chosen, the conformal Ricci scalar is zero, $\bar{R}=0$.
The remaining metric term in the Hamiltonian constraint is the $\nabla^{2}e^{\phi}$
term, fluctuations of which will correspond to fluctuations in $\rho$.
In order to produce density fluctuations described by the above cosmologically-motivated
power spectrum, we choose the conformal factor $\phi$ to be described
by a related power spectrum,
\begin{equation}
P_{\phi\phi}=k^{-4}P_{\delta\delta}\,.
\end{equation}
We generate a Gaussian random realization of $\phi$ according to
this prescription. The remaining metric variable $K$, the local expansion
rate, is chosen to correspond to a desired Hubble expansion rate.
The density is then fixed by the Hamiltonian constraint equation. Further
details on this method can be found in \cite{1511.01106}.

Although the initial conditions we use are qualitatively similar to
those found in a cosmological setting, the setup we use does not precisely
correspond to physical expectations. We therefore recognize this as
a toy model, rather than a precision calculation. Nevertheless, we
are hopeful that the effects we see here can provide a reliable indication
of the the order of magnitude of corrections to Newtonian calculations
due to general relativistic effects, and that they can provide a qualitative
indication of the relevance of relativistic effects to observations.
Generalizing these initial conditions to more closely correspond to
expectations from Newtonian or linear theory, but in a relativistic
setting, will be an important future task.

The final ingredient required in order to specify the sky seen by
an observer is, of course, an observer. In this work, we lay down
initial conditions and integrate the simulation forward to a desired
time of observation. We then place an observer at the center of our
simulation volume and integrate along geodesics away from this observer
in {\sc Healpix} \cite{astro-ph/0409513} directions,
from the observer's spacetime point ``backwards'' in time. This
observer, along with sources, are taken to be at rest in geodesic slicing,
or to be co-moving with the local fluid.

\section{Results}
\label{sec:Results}

Here we present results from a simulation in which photon geodesics are integrated
from an observer back in time to a redshift of $z = 0.25$. In particular, we compare 
Eq.~\ref{eq:gr_convergence} to Eq.~\ref{eq:linear_convergence_expr}. In order
to compute the former of these we utilize the above 3+1 formulation of
Einstein's equations, thus integrating through a fully general relativistic
spacetime including no approximations or reductions to the Einstein field
equations. The latter of these expressions originates from a linearized, scalar
gravity treatment, for which we additionally utilize the Born approximation in
order to obtain a radial coordinate as described in Sec.~\ref{subsec:convergence}.

\begin{figure}[htb]
  \centering
    \includegraphics[width=0.45\textwidth]{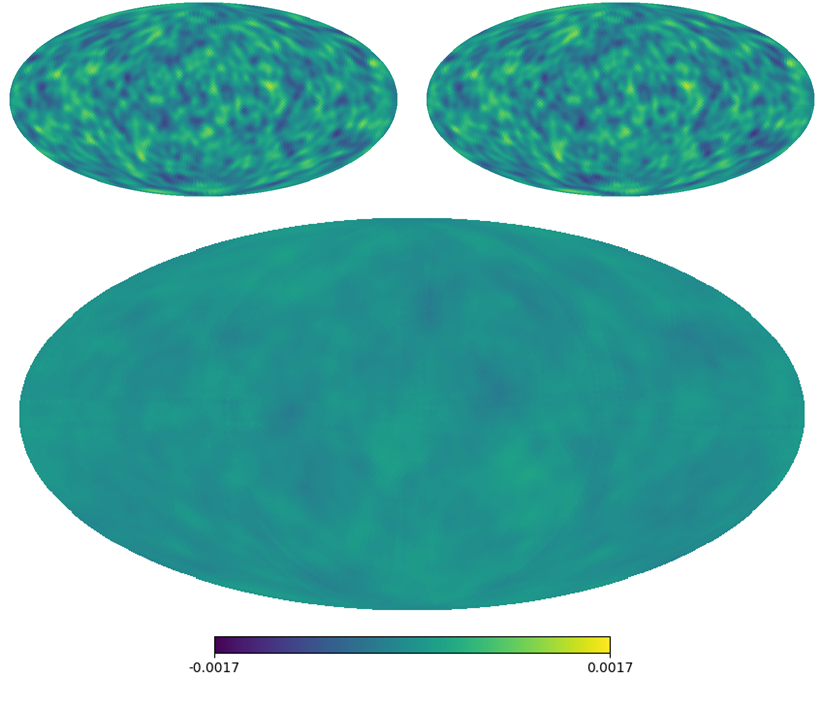}
    \caption{Observed skies: top left is a sky generated using approximate theory (Eq.~\ref{eq:linear_convergence_expr}),
      top right a relativistic sky (Eq.~\ref{eq:gr_convergence}), and bottom the difference between the two. These
      skies are generated using a {\sc Healpix} resolution of Nside = 32, and all
      maps have had the angular monopole and dipole contributions removed.}
  \label{Healpix_grf_sky}
\end{figure}

\begin{figure}[htb]
  \centering
    \includegraphics[width=0.5\textwidth]{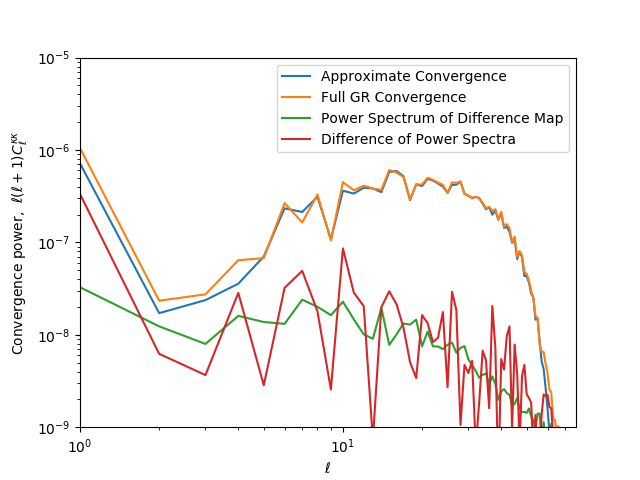}
    \caption{Power spectra of the simulated skies shown in Figure~\ref{Healpix_grf_sky}.
    The orange curve depicts results from a fully relativistic run, blue from
    approximate theory, and green the power spectrum of the difference
    of the difference map.
    }
  \label{convergence_grf_ps}
\end{figure}

In order to compute the angular power spectrum, we decompose the convergence field
on the sky into spherical harmonics. The convergence field is written as
\begin{equation}
\kappa(\theta,\phi)=\sum_{lm}a_{lm}Y_{lm}(\theta,\phi)\,.
\end{equation}
From this, the angular power spectrum is then defined,
\begin{equation}
C_{l}^{\kappa\kappa}=\frac{1}{2l+1}\sum|a_{lm}^{2}|\,.
\end{equation}
In practice, we compute the angular power spectrum by integrating angular
diameter distances---and therefore convergences---in {\sc Healpix} directions
for an observer in our simulated universe. We then use standard {\sc Healpix}
routines to compute the power spectrum from the convergence maps we produce.
Convergence maps are plotted in Figure~\ref{Healpix_grf_sky}, depicting the
difference between relativistic simulation results and approximate results.
The power spectra that correspond to these images are shown in
Figure~\ref{convergence_grf_ps}.

In order to obtain meaningful results, we must also compute the numerical
error for convergence values along each geodesic. We do so by performing runs
using a set of four resolutions in our simulation, $N^{3}=128^{3}$, $160^{3}$,
$192^{3}$, and $256^{3}$. We then Richardson extrapolate continuum limit
convergence values using different pairs of runs, and use the distribution of
extrapolated values to provide us with a measure of uncertainty in these
convergence values. The values typically agree at one part in $10^{4}$, or at
a level significantly smaller than the difference between convergences computed
using approximate and relativistic methods. The uncertainty in extrapolated power
spectra is also found to be accurate at this level. As an additional note, we
compute power spectra using $\ell_{\rm max}\sim2.4{\rm max}(\ell)$ in {\sc Healpix}
in order to obtain more accurate results. The resulting numerical error in the
spectra we present in this paper is then expected to be better than a
part in $10^{4}$.

\begin{figure}[htb]
  \label{convergence_grf_ps_combined}
  \centering
    \includegraphics[width=0.5\textwidth]{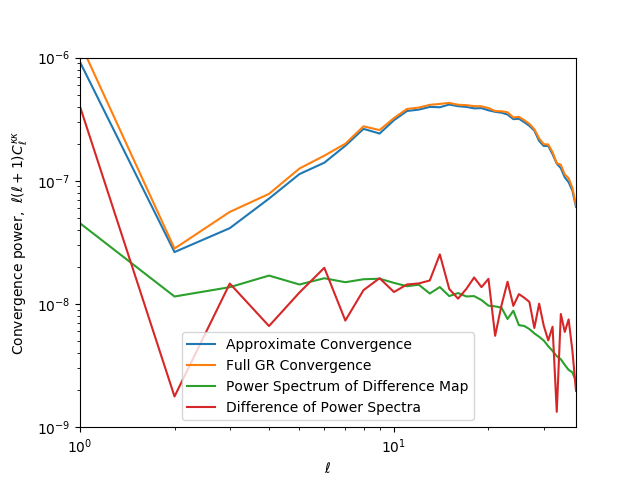}
    \caption{Shown are averages of 20 power spectra obtained from independent
    simulations. The orange curve shows the fully relativistic power spectrum,
    and blue the approximate spectrum. The green curve is the average power
    spectrum of the difference maps, and the red power spectrum the average
    of a direct subtraction of the GR and approximate power spectra.}
\end{figure}

There is, in addition, sampling error---or cosmic variance---resulting from the
limited number of simulations we run. In order to address this, we simulate
twenty skies in total, and average the power spectra together. The resulting
spectra are shown in Figure~\ref{convergence_grf_ps_combined}, in which we find
an $\ell$-dependent increase in the amplitude of the approximate power spectrum
compared to the fully relativistic spectrum.
Some remaining cosmic variance can be seen as ripples or wiggles in the power
spectra; such effects may be expected to diminish as an increasing number of
simulations are averaged over.

Finally, we remark on the potential origins of the discrepancies we see: are
these due to nonlinear physics, the Born approximation, or merely artifacts of
the gauge transformations we have performed? The amplitude of the Newtonian
potentials, and amplitude of the components of the gauge transformation are
not large, being of order a part in $10^5$ on these large scales. Fluctuations in
the synchronous gauge metric itself, $\sigma_\phi / \phi$, are closer to a part
in $10^4$. One may therefore expect ambiguities due to gauge and nonlinear effects
to be smaller than the observed percent-level corrections, indicating physical
approximations made may be breaking down. However, we also find that the
Newtonian potentials $\Phi$ and $\Psi$ evolve towards percent-level disagreement,
or that a significant gravitational slip develops, suggesting the system
is evolving away from the linearized constraint equations typically enforced in a
cosmological setting. Further exploration will be required to precisely
characterize the physics at play here, and to determine how both physical and
perturbative approximations are breaking down.

\section{Discussion}

In this manuscript, we have described the possibility of percent-level
corrections to lensing power spectrum predictions due to a fully relativistic
treatment of gravitational lensing by large-scale structure. This suggests
circumspection in the utilization of weak lensing measurements to constrain
cosmological parameters. However, a direct comparison of our work to prior
literature on lensing cosmology is not possible at the present time. Due to
computational limitations, we work within a toy, inhomogeneous Einstein-de Sitter
cosmology, explore only large angular scales ($\ell \sim 10$), and only consider
lensing out to a redshift of $z \sim 0.25$. By way of contrast, lensing by ongoing
and forthcoming observational facilities is dominated by structure at significantly
higher redshifts ($z \sim 0.6-1$), and the majority of the cosmological information
is contained in lensing correlations on considerably smaller scales (significantly
less than a degree, multipoles of $\ell \gtrsim 300$).

We also do not currently have a reliable method of extrapolating our results to the
more practical case of small-scale lensing correlations induced by high-redshift,
large-scale structure in a dark energy-dominated universe. However, it is
interesting to speculate on the possible importance of our work in this context.
Using the methods of \cite{1111.0052} and \cite{1405.6205} it is straightforward
to estimate the potential impact of the systematic errors that we explore on the
program to constrain cosmology using weak gravitational lensing correlations.
For an LSST- or Euclid-like survey, we estimate that a one-percent systematic
offset in the lensing power spectrum corresponds to a systematic error on the
inferred dark energy equation of state parameter, $w$, that is roughly twice
the statistical error with which this parameter may be measured. We estimate a
similar level of error for the neutrino mass. We argue that this is strong
motivation to pursue fully relativistic lensing studies further. However, we
emphasize that these estimates remain speculative as we do not yet
understand the cosmology dependence, scale dependence, or redshift dependence
of the effects we describe, and all of those factors can significantly alter
these estimates.

Future studies may also wish to examine the behavior of specific dark energy or
dark matter models in a fully relativistic context. Important effects have been
considered using approximate treatments in the past, including baryonic physics
\cite{1501.02055}, radiation \cite{1703.08585}, interactions of propagating light
with contents of the Universe \cite{1705.02328}, and a more complete picture of
phase space dynamics \cite{1509.01699}. Incorporating such phenomenology into
a fully general relativistic simulation has not yet been performed in a
cosmological setting, and will be an important task for relativistic simulations
in the coming years.

\section{Acknowledgments}
We would like to thank Marcio O'Dwyer for valuable conversations. JTG is
supported by the National Science Foundation, PHY-1414479; JBM and GDS are
supported by a Department of Energy grant DE-SC0009946 to CWRU; and
ARZ is supported in part by the DoE through Grant DE-SC0007914 and by 
the Pittsburgh Particle physics Astrophysics and Cosmology Center (Pitt PACC) 
at the University of Pittsburgh. The simulations
in this work made use of the High Performance Computing Resource in the Core
Facility for Advanced Research Computing at Case Western Reserve University,
and of hardware provided by the National Science Foundation and the Kenyon
College Department of Physics.

\bibliography{references}

\end{document}